\documentclass[aps,%
12pt,%
oneside,
onecolumn,%
musixtex, %
superscriptaddress,%
centertags]{article} %%
\usepackage[utf8x]{inputenc}
\usepackage[noend]{algorithmic}
\usepackage[ruled]{algorithm}
\usepackage{braket}
\usepackage{amsmath}
\usepackage{comment}
\usepackage{setspace}
\usepackage{titlesec}
\usepackage{setspace}

\usepackage{color}
\usepackage{amsthm}

\usepackage[english]{babel}

\usepackage[T2A]{fontenc} % Поддержка русских букв
\usepackage[colorlinks=true,linkcolor=blue,unicode=true]{hyperref}
\usepackage{euscript}
\usepackage{supertabular}
\usepackage[pdftex]{graphicx}%
\usepackage{amsthm,amssymb, amsmath}
\usepackage{textcomp}
\usepackage[noend]{algorithmic}
\usepackage[ruled]{algorithm}
\usepackage{braket}
\usepackage{amsmath}
\usepackage{comment}
\usepackage{setspace}
\DeclareMathOperator{\Tr}{Tr}
\usepackage{titlesec}
\usepackage{setspace}

\usepackage{color}
\usepackage[dvipsnames]{xcolor}

\newcommand{\be}{\begin{equation}}
\newcommand{\ee}{\end{equation}}
\newcommand{\ba}{\begin{eqnarray}}
\newcommand{\ea}{\end{eqnarray}}
\newcommand{\bc}{\begin{comment}}
\newcommand{\ec}{\end{comment}}
\newcommand{\eps}{\varepsilon}
\newcommand{\bs}{\boldsymbol}
\newcommand{\on}{\operatorname}
\newcommand{\nn}{\nonumber}

\definecolor{mygray}{gray}{0.7}

\usepackage{amsthm}

\usepackage[english]{babel}

\begin{document}

%%%%%%%%%%%%%%%%%%%%%%%%%%%%%%%%%%%%%%%%%%

\begin{center}

{\Large\bf The Franke-Gorini-Kossakowski-Lindblad-}\\
{\Large\bf Sudarshan (FGKLS) Equation for Two-Dimensional Systems}

\vspace{1cm}
{\large Alexander A. Andrianov$^{1,2,}$\footnote{E-mail: a.andrianov@spbu.ru}, Mikhail V. Iof\/fe$^{1 ,}$\footnote{E-mail: m.ioffe@spbu.ru}, \\
Ekaterina A. Izotova$^{1,3,4,}$\footnote{E-mail: ekat.izotova@gmail.com}, Oleg O. Novikov$^{1 ,}$\footnote{E-mail: o.novikov@spbu.ru}}
\\
\vspace{0.5cm}
$^1$ Department of High Energy and Elementary Particle Physics, Faculty of Physics, Saint Petersburg State
University, 7/9 Universitetskaya nab., St. Petersburg 199034, Russia \\
$^2$ Departament de Física Quàntica i Astrofìsica and Institut de Ciències del Cosmos (ICCUB), Universitat de
Barcelona, Martí i Franquès 1, 08028 Barcelona, Spain \\
$^3$ Center for Photonic Science and Engineering, Skolkovo Institute of Science and Technology,
Bolshoy Boulevard 30, bld. 1, Moscow 121205, Russia \\
$^4$ Department of Theoretical Astrophysics and Quantum Field Theory, Phystech School of Fundamental and
Applied Physics, Moscow Institute of Physics and Technology, Institutsky Lane 9,
Dolgoprudny 141700, Russia 
\end{center}

\vspace{1cm}

\begin{abstract}
Open quantum systems are, in general, described by a density matrix that is evolving under transformations belonging to a dynamical semigroup. They can obey the Franke--Gorini--Kossakowski--Lindblad--Sudarshan (FGKLS) equation. We exhaustively study the case of a Hilbert space of dimension $2$. First, we find final fixed states (called pointers) of an evolution of an open system, and we then obtain a general solution to the FGKLS equation and confirm that it converges to a pointer. After this, we check that the solution has physical meaning, i.e., it is Hermitian, positive and has trace equal to $1$, and find a moment of time starting from which the FGKLS equation can be used---the range of applicability of the semigroup symmetry. Next, we study the behavior of a solution for a weak interaction with an environment and make a distinction between interacting and non-interacting cases. Finally, we prove that there cannot exist oscillating solutions to the FGKLS equation, which would resemble the behavior of a closed quantum system.
\end{abstract}


\section{Introduction}

Closed quantum systems are described by the Schrödinger equation. If we take a pure initial state, it will always stay pure during the evolution of the system undergoing the unitary symmetry transformation. This equation can be put in another form, using a density matrix---it is a Liouville equation.

When we proceed to open quantum systems, we find that environmental influences may transform an initial pure-state density matrix into the mixed-state density matrix \cite{breuer, schlosshauer, weiss}. Then, the Schrödinger equation can no longer describe this transition; a density matrix is evolving under transformations belonging to a dynamical semigroup and the  standard equation for this case is the Franke--Gorini--Kossakowski--Lindblad--Sudarshan (FGKLS) equation~\cite{franke, gorini, lindblad} :

\be
\label{lindblad}\dot{\rho} = -i[H,\rho]+\sum_a L^{(a)}  \rho L^{(a),\dag} - \frac{1}{2} \left\{\sum_a  L^{(a),\dag} L^{(a)},\rho\right\}
\ee
{Here,} %MDPI:  can add indentation? please confirm the following text after equation without noindnetation, please confirm and revise.
 $\rho$ is a density matrix of the system, $H$ is a Hamiltonian of that system, $ L^{(a)}$ are a set of operators that incorporate an interaction of a system with an environment. The latter Lindbladians generate deviations from purely unitary evolution and the entire evolution operators belong to a set of a dynamical semigroup.   This approach has quite a few applications in the analysis of decoherence in nonrelativistic quantum systems \cite{breuer, schlosshauer, weiss}. We notice also that recently the FGKLS approach has been found to be fruitful in high-energy \mbox{physics \cite{akamatsu, Blaizot, hep}}, \protect condensed matter \cite{cond1,cond3,cond4} and quantum biology \cite{bio2,bio3}.

This equation indeed describes well
open quantum systems. We can see this if we take the system
of interest and an environment as a large closed system that is described by
the Liouville equation. Then, provided that the current understanding of the quantum
theory remains to be valid on the fundamental level, the FGKLS equation
should arise as an effective description of a subsystem of the larger system
that includes environmental degrees of freedom that is assumed to undergo
the unitary evolution driven by a \mbox{self-adjoint Hamiltonian.}

It will be of the FGKLS form if one ensures the positivity and completeness (its trace equal to $1$) of the density matrix, as well as its complete positivity \cite{benatti_floreanini}, which delivers exactly this form of the equation. It is linear and supports the superposition of quantum states. Thus, it may substitute the Schrodinger dynamics of wave functions  by the extended fundamental dynamics of density matrices \cite{ weinberg_2014}. The Lindblad operators $L^{(a)}$ may be associated with elements of measuring apparatus \cite{weinberg_2014} or with an external noise  \cite{andrianovtarrach}. This kind of quantum dynamics may be referred to as a non-Hamiltonian one \cite{tarasov}.

As the system interacts with an environment,   the process of \textit{{decoherence} %MDPI:  is the italic necessary?.
} takes over. During this process, pure quantum states transform into the mixed ones, the information about the initial state partially becomes lost, and the system becomes more classical. As expected, the system  may reach a certain final state, which is the same for all initial states. Such final states are called \textit{{pointers}%MDPI: is the italic necessary?.
} \cite{zurek}. The fact that they are robust and do not change over time is expressed by the following simple equation:
\be
\label{statpointer} \dot\rho = 0,\quad t \geq t_{final}.
\ee

The goal of this work is to study in detail the decoherence process for the systems of a small dimension of the Hilbert space equal to $2$. We exhaustively consider all possible forms of a Lindblad operator: diagonal and Jordan block type. In the second section, we find pointers for these two cases. In the third section, we obtain a general solution to the FGKLS equation and confirm that it indeed converges to a pointer with time. In the fourth section, we check that the obtained solution has physical meaning, i.e., the density matrix is Hermitian, positive and has trace equal to $1$. In the fifth section, we explore how the solution behaves for weak interaction with an environment. In the sixth section, we make conclusions about the existence of oscillation solutions, resembling the solutions of a closed system. In the seventh section, we summarize and make final remarks. This work is partially based on our previous study in \cite{pertalg}.
\section{Pointers for the FGKLS Equation in Two Dimensions}\label{sec2}

Let us consider the FGKLS equation for the systems in two-dimensional Hilbert space: both the Hamiltonian $H$ and the Lindblad operator ({For simplicity,} %MDPI: Footnote is not permitted in this journal, so we have moved it into the text, please confirm the whole text..
 the case of one Lindblad operator $L$ will be considered. Generalization to the systems with many operators $L^{(a)}$ is straightforward.) $L$ describing interaction with the environment are $2 \times 2$ operators. We build the Hilbert space spanned on a two-level energy orthonormal basis of\linebreak $H \ket{E_l}=E_l\ket{E_l}; \, l=1,2$: an arbitrary orthonormal basis in this space is $\ket{\psi_i} = \sum_{k=1}^{k=2}u_{ik} \ket{E_k} $ with unitary matrix $U$ coefficients.
The Hamiltonian, generated by a Hermitian matrix of the size $2\times 2$, takes the form
\be \label{h2} H= \sum_{i,k=1}^{i,k=2}\eps_{ik}\ket{\psi_i} \bra{\psi_k} .
\ee
{On} %MDPI: can add indentation? please confirm the following text after equation without noindnetation, please confirm and revise.
 the same basis, the Lindblad operator is defined as:
\be \label{l2} L = c \sum_{i,k=1}^{i,k=2}l_{ik} \ket{\psi_i} \bra{\psi_k}
\ee
with arbitrary complex coefficients $l_{ik}$ and real coefficient $c$, which will be useful further on to control the behavior of solutions for small $L.$

The density matrix
\be \label{rho2} \rho = \sum_{i,k=1}^{i,k=2}f_{ik}(t)\ket{\psi_i} \bra{\psi_k}
\ee
has to satisfy the well-known FGKLS equation
\be
\label{lindblad1}\dot{\rho} = -i[H,\rho]+ L \rho L^\dagger - \frac{1}{2} \left\{ L^\dagger L,\, \rho\right\}
\ee
and to obey the following properties:

1. $\rho$ is Hermitian: $\rho=\rho^\dag$, i.e.,
\be \label{herm2}  f_{11}, f_{22} \in \mathbb{R}; \;\;\; f_{21} = f_{12}^* \ee

{2.} $\Tr \rho = 1$:  %add indnentation, please confirm
\be \label{trace2} f_{11} + f_{22} = 1 \ee

{3.} $\rho$ is non-negative.

Further on, the first two properties will be taken into account directly in the equations, while the last one will be checked
for the obtained solutions of (\ref{lindblad}) postfactum. By default, the basis elements $|\psi_i >$ as well as coefficients $\eps_{ik}$ and $l_{ik}$ are assumed to be time-independent.

Substituting the expressions (\ref{h2})--(\ref{rho2})  into the FGKLS equation (\ref{lindblad}) and taking into account the first two properties of $\rho$ above, we {obtain:} %MDPI: is the bold format in equation (f11, f12,f21,f22...) necessary? If not, please remove bolds. And does the asterisk need explanations?
\be
\boldsymbol{\dot f_{11}} = A \boldsymbol{f_{11}} + B \boldsymbol{f_{22}} + E  \boldsymbol{f_{12}} + E^*  \boldsymbol{f_{21}}
%\nonumber
\ee
\be
%\label{f22}
\boldsymbol{\dot f_{22}} = -A \boldsymbol{f_{11}} - B \boldsymbol{f_{22}} -  E  \boldsymbol{f_{12}} - E^*  \boldsymbol{f_{21}}
%\nonumber
\ee
\be
%\label{f12}
\boldsymbol{\dot f_{12}} = G \boldsymbol{f_{11}} + H \boldsymbol{f_{22}} + J  \boldsymbol{f_{12}} + K  \boldsymbol{f_{21}}
%\nonumber
\ee
\be
%\label{f21}
\boldsymbol{\dot f_{21}} = G^* \boldsymbol{f_{11}} + H^* \boldsymbol{f_{22}} + K^*  \boldsymbol{f_{12}} + J^*  \boldsymbol{f_{21}},
%\nonumber
\ee
where
\be \label{bvar} B = c^2 |l_{12}|^2  \ee
\be \label{evar} E = i\eps_{21} + \frac{1}{2} c^2 (l_{11} l^*_{12} - l^*_{22} l_{21}) \ee
\be \label{gvar} G = i\eps_{12} + c^2( l_{11} l^*_{21} - \frac{1}{2}l^*_{11} l_{12} - \frac{1}{2} l^*_{21}l_{22}) \ee
\be \label{hvar} H = -i\eps_{12} + c^2( l^*_{22} l_{12} - \frac{1}{2} l^*_{11} l_{12} - \frac{1}{2} l^*_{21} l_{22} ) \ee
%вырождение обозначений
\be \label{jvar} J = -i \Delta\eps + c^2( l_{11} l^*_{22} - \frac{1}{2} |l_{11}|^2 - \frac{1}{2} |l_{22}|^2 - \frac{1}{2} |l_{12}|^2 - \frac{1}{2} |l_{21}|^2) \ee
\be \label{kvar} K = c^2  l_{12} l^*_{21} \ee
where $\Delta\eps \equiv \eps_{11} - \eps_{22}$.

By definition, the pointers of the FGKLS equation are solutions $\rho^{(p)}$ of (\ref{lindblad}), which become stable asymptotically for large $t \to\infty$, i.e., $\dot{f}^{(p)}_{ik}(t\to\infty)=0.$
Thus, we have the system of three linear equations for independent variables $f^{(p)}_{11}, f^{(p)}_{12}, f^{(p)\,\star}_{12}:$
\be (A-B) \bs{f^{(p)}_{11}} + E \bs{f^{(p)}_{12}} + E^* \bs{f^{(p)\,\star}_{12}} = -B
%\nonumber
\ee
\be (G-H) \bs{f^{(p)}_{11}} + J \bs{f^{(p)}_{12}} + K \bs{f^{(p)\,\star}_{12}} = -H
%\nonumber
\ee
\be (G^* - H^*) \bs{f^{(p)}_{11}} + K^* \bs{f^{(p)}_{12}} + J^* \bs{f^{(p)\,\star}_{12}} = -H^*
%\nonumber
\ee
or, in a matrix form:
\ba
&&\!\!\!\!\!\!\!\!\!\!\!\!\begin{pmatrix}
- |l_{12}|^2 - |l_{21}|^2 & i e_{21} + \frac{1}{2} (l_{11}l^*_{12} - l^*_{22} l_{21}) & -ie_{12} +  \frac{1}{2} (l^*_{11} l_{12} - l_{22} l^*_{21}) \\
2ie_{12} + l_{11} l^*_{21} - l^*_{22} l_{12} & \frac{J}{c^2} & l_{12}l^*_{21} \\
-2ie_{21} + l^*_{11} l_{21} - l_{22} l^*_{12} & l^*_{12}l_{21} & \frac{J^*}{c^2} \\
\end{pmatrix} \nonumber \\
&& \times\begin{pmatrix}
\bs{f^{(p)}_{11}} \\
\bs{f^{(p)}_{12}} \\
\bs{f^{(p)\,\star}_{12}}\\
\end{pmatrix} = \begin{pmatrix}
-|l_{12}|^2 \\
- \frac{H}{c^2}\\
- \frac{H^*}{c^2} \\
\end{pmatrix},   \label{matreq1}
\ea
where $J$ and $H$ are given {by} %MDPI: can change to (\ref{hvar}) and (\ref{jvar})? please confirm
 (\ref{jvar}) and (\ref{hvar}),
 and $e_{ij}$ are defined as $e_{ij} \equiv \frac{\eps_{ij}}{c^2}$.
The existence of pointers depends on the determinant of the matrix in (\ref{matreq1}). After choosing an appropriate basis, a Lindblad operator $L$ can be reduced to one of two possible forms:

\begin{itemize}
\item The Lindblad operator is diagonal;

\item The Lindblad operator is a Jordan block.
\end{itemize}

%%%%%%%%%%%%%%%%%%%%%%%%%%%%%%%%%%%%%%%%%%
\subsection{Diagonal Lindblad Operator}

For the diagonal operators $L$,
\be \label{ldiag} L = c \begin{pmatrix}
\lambda_1 & 0 \\
0 & \lambda_2\\
\end{pmatrix}
\ee
with complex values $\lambda_i,$ Equation (\ref{matreq1}) for the pointers becomes:
\be \label{matreq1abdiag}
\begin{pmatrix}
0 & a & a^* \\
-2 a^* & b & 0 \\
-2a & 0 & b^* \\
\end{pmatrix}\times
\begin{pmatrix}
\bs{f^{(p)}_{11}} \\
\bs{f^{(p)}_{12}} \\
\bs{f^{(p)\,\star}_{12}} \\
\end{pmatrix}
=
\begin{pmatrix}
0 \\
-a^*\\
-a \\
\end{pmatrix}
\ee
where $a = i e_{21}$ and $b = -i (e_{11} - e_{22}) + \lambda_1 \lambda^*_2 - \frac{1}{2} |\lambda_1|^2 - \frac{1}{2} |\lambda_2|^2$.
The determinant of the matrix in (\ref{matreq1abdiag}) is $2|a|^2 (b+b^*)={-2|e_{21}|^2 |\lambda_1 - \lambda_2|^2},$
and several cases have to be considered:
\begin{itemize}
\item Determinant of the matrix in (\ref{matreq1abdiag}) does not vanish:
$$|e_{21}|^2 |\lambda_1 - \lambda_2|^2 \neq 0,$$
and the solution is:
\be
%\label{pointerdiag}
\rho^{(p)} =
\begin{pmatrix}
\frac{1}{2} & 0 \\
0 & \frac{1}{2}\\
\end{pmatrix}.
%\nonumber
\ee
{This} pointer does not depend on the parameters $c,\lambda_1,\lambda_2$ in $L,$ being a maximally mixed state.
\item Determinant vanishes due to $e_{21} = 0.$ Two options in this case exist:

\begin{enumerate}
\item If, additionally, $b$ vanishes, i.e., $-i (e_{11} - e_{22}) + \lambda_1 \lambda^*_2 - \frac{1}{2} |\lambda_1|^2 - \frac{1}{2} |\lambda_2|^2 = 0,$
the pointer is an arbitrary matrix with unit trace:
\be \label{pointerdiag1}
\rho^{(p)} =
\begin{pmatrix}
f^{(p)}_{11} & f^{(p)}_{12} \\
f^{(p)}_{21} & 1 - f^{(p)}_{11}\\
\end{pmatrix}.
\ee

\item If, additionally, $b$ does not vanish, i.e., $-i (e_{11} - e_{22}) + \lambda_1 \lambda^*_2 - \frac{1}{2} |\lambda_1|^2 - \frac{1}{2} |\lambda_2|^2 \neq 0,$ the pointer is an arbitrary diagonal matrix with unit trace:
\be \label{pointerdiag2}
\rho^{(p)} =
\begin{pmatrix}
f^{(p)}_{11} & 0 \\
0 & 1 - f^{(p)}_{11}\\
\end{pmatrix}
\ee

\end{enumerate}

\item Determinant vanishes due to $\lambda_1 = \lambda_2.$ Then,
the pointer is expressed via one arbitrary real parameter $x\equiv f^{(p)}_{11}:$
\be
%\label{pointerdiag3}
\rho^{(p)} =
\begin{pmatrix}
x & e_{12} \frac{2x-1}{e_{11}-e_{22}} \\
e_{21} \frac{2x-1}{e_{11}-e_{22}} & 1 - x\\
\end{pmatrix}.
%\nonumber
\ee

\end{itemize}

\subsection{Lindblad Operator of a Jordan Block Form}

Let us consider the operator $L$ of the Jordan block form:
\be \label{ljordan} L = c(\lambda \hat I + \sigma_+ ) =
c \begin{pmatrix}
\lambda & 1 \\
0 & \lambda\\
\end{pmatrix};
\ee
{with complex $\lambda_i$ and real parameter $c,$ which will allow us to make operator $L$ small} \mbox{if necessary.}

A remark on the form (\ref{ljordan}) is appropriate here. The FGKLS equation with such $L, \, L^{\dag}$ can be transformed as follows:
\ba
\dot \rho &=& - i[H, \rho] + c^2( \lambda\hat I + \sigma_+) \rho ( \lambda^{\star}\hat I + \sigma_-)- \nonumber \\
& & \frac12 c^2 \rho( \lambda^{\star}\hat I + \sigma_-) ( \lambda\hat I + \sigma_+) - \frac12 c^2( \lambda^{\star}\hat I + \sigma_-) ( \lambda\hat I + \sigma_+)\rho \nonumber  \\
&=& - i[\widetilde H, \rho] + c^2 \sigma_+\rho \sigma_- -\frac12 c^2( \sigma_- \sigma_+\rho + \rho\sigma_- \sigma_+).
%\nonumber
\ea
{This} means the invariance of our model under simultaneous replacements:
\ba
L = c ( \lambda\hat I + \sigma_+) &\to & \widetilde L = c\sigma_+, \label{r1}\\
H &\to & \widetilde H = H - \frac{ic^2}{2}(\lambda\sigma_- - \lambda^{\star}\sigma_+). \label{r2}
\ea
{Thus}, instead of the Jordan block (\ref{ljordan}) form of the Lindblad operator, the nilpotent operators $\sigma_{\pm}$ might be used up to a shift in the Hamiltonian onto a fixed Hermitian matrix. These two models are equivalent, and below, the form (\ref{ljordan}) will be used.

A possible simulation of environmental effects to obtain a Lindblad operator of a Jordan block form is presented in the example in Appendix \ref{appc}.

Now, we come back to solving the FGKLS equation for a general case of a Lindlad operator of a Jordan block form.

Matrix equation (\ref{matreq1}) is simplified as:
\be \label{matreq1ab}
\begin{pmatrix}
-1 & a & a^* \\
-2a^* & b & 0 \\
-2a & 0 & b^* \\
\end{pmatrix}
\begin{pmatrix}
\bs{f^{(p)}_{11}} \\
\bs{f^{(p)}_{12}} \\
\bs{f^{(p)\star}_{12}} \\
\end{pmatrix}
=
\begin{pmatrix}
-1 \\
-a^* \\
-a \\
\end{pmatrix}
\ee
where $a = ie_{21} + \frac{1}{2} \lambda$ and $b = -\frac{1}{2} - i \Delta e$, $\Delta e \equiv e_{11}-e_{22}$.

The expression for the determinant of the matrix in (\ref{matreq1ab}) is obviously negative:
$${-|b|^2 + 2|a|^2(b+b^*)} = {= - \frac{1}{4} - (\Delta e)^2 - 2|e_{21}-\frac{1}{2}i\lambda|^2  < 0},$$
and the solution of Equation (\ref{matreq1ab}) $\{f^{(p)}_{11}, f^{(p)}_{12}, f^*_{p12}\}$ exists:
\ba
&&\rho_p =\begin{pmatrix}
f^{(p)}_{11} & f^{(p)}_{12} \\
f^{(p)}_{21} & f^{(p)}_{22} \\
\end{pmatrix} =
\frac{1}{2 |a|^2 + |b|^2}
\begin{pmatrix}
|a|^2+|b|^2 & a^* b^* \\
ab & |a|^2\\
\end{pmatrix} = \nonumber \\
&&= \gamma \begin{pmatrix}
 \frac{1}{4} + (\Delta e)^2 + |e_{21}+\frac{1}{2}i\lambda|^2 & (- ie_{12} + \frac{1}{2} \lambda^*)(-\frac{1}{2} + i \Delta e ) \\
( ie_{21} + \frac{1}{2} \lambda)(-\frac{1}{2} - i \Delta e ) & |e_{21}-\frac{1}{2}i\lambda|^2 \\
\end{pmatrix}, \label{pointer2}\\
&&\gamma = \frac{1}{ \frac{1}{4} +(\Delta e)^2 + 2|e_{21}-\frac{1}{2}i\lambda|^2}
\ea
{This} solution is physically reasonable, since $\Tr \rho_p = 1$, and positivity of $\rho_p$ is provided by inequality $\det \rho_p > 0$.
\begin{itemize}
\item For the special case of degenerate $H,$ then $\eps_1 = \eps_2$, $H = \eps \begin{pmatrix}
1 & 0 \\
0 & 1 \\
\end{pmatrix}$, the pointer is
\be
\rho_{p, \text{deg}} =
\frac{1}{ 2|\lambda|^2 + 1}
\begin{pmatrix}
|\lambda|^2 + 1 & -  \lambda^*\\
- \lambda &  |\lambda|^2  \\
\end{pmatrix}
\ee
{It} does not depend on parameter $c$. This is because, for such $H$, the first term in the Lindblad equation (\ref{lindblad}) with pointer condition (\ref{statpointer}) disappears, and the parameter $c$, after substituting $L$ (\ref{ljordan}), falls away.

\end{itemize}

In Appendix \ref{appa}, one can find the expression for a pointer for weak interaction \mbox{with the environment}.

\section{The General Solution of the FGKLS Equation in Two-Dimensional Hilbert Space}\label{sec3}

\subsection{The Case of Diagonal Lindblad Operator}

To find the general solution of the FGKLS equation (\ref{lindblad}) with the Lindblad operator of diagonal form (\ref{ldiag}), one has to solve the system of three linear equations:
\be \bs{\dot f_{11}} =  i \eps_{21} \bs{f_{12}} -i\eps_{12} \bs{f_{21}}
%\nonumber
\ee
\be \bs{\dot f_{12}} = 2i \eps_{12} \bs{f_{11}} + \left( -i \Delta\eps  - \frac{1}{2} c^2|\lambda_1|^2 - \frac{1}{2} c^2|\lambda_2|^2 + c^2\lambda_1 \lambda^*_2 \right) \bs{f_{12}} -i\eps_{12}
%\nonumber
\ee
\be \bs{\dot f_{21}} = -2i \eps_{21} \bs{f_{11}} + \left( i \Delta\eps  - \frac{1}{2} c^2|\lambda_1|^2 - \frac{1}{2} c^2|\lambda_2|^2 + c^2\lambda^*_1 \lambda_2 \right) \bs{f_{21}} +i\eps_{21}
%\nonumber
\ee
{The} trace condition (\ref{trace2}) will give the solution for the variable $f_{22}$ as well.
The solution of this system of differential equations is the sum of a partial solution of the non-homogeneous system and the general solution of the homogeneous system.

Since only the pointer (\ref{pointer2}) can obviously play a role of partial solution to this system,
the problem left is to find the general solution of the homogeneous system of equations, which in the matrix form is:
\ba
&&\begin{pmatrix}
\dot f_{11} \\
\dot f_{12} \\
\dot f_{21} \\
\end{pmatrix} = \nonumber\\
&&\!\!\!\!\!\!\!\!\!\!\!\!\!\!\begin{pmatrix}
0 & i \eps_{21} & -i \eps_{12} \\
2i \eps_{12}&-i \Delta\eps  - \frac{1}{2} c^2|\lambda_1|^2 - \frac{1}{2} c^2|\lambda_2|^2 + c^2\lambda_1 \lambda^*_2 & 0 \\
-2i \eps_{21}& 0 & \!\!\!\!\!\!\!\!\!\!i \Delta\eps  - \frac{1}{2} c^2|\lambda_1|^2 - \frac{1}{2} c^2|\lambda_2|^2 + c^2\lambda^*_1 \lambda_2 \\
\end{pmatrix} \nonumber\\
&&\times\begin{pmatrix}
f_{11} \\
f_{12} \\
f_{21} \\
\end{pmatrix}
\equiv A \begin{pmatrix}
f_{11} \\
f_{12} \\
f_{21} \\
\end{pmatrix}\label{homdiffeq}
\ea

We will look for solutions in the form
\be \label{x} \begin{pmatrix}
f_{11} (t) \\
f_{12} (t) \\
f_{21} (t) \\
\end{pmatrix}
 \equiv \vec X(t) = e^{\Lambda t} \vec V \ee
where $\vec V$ is a constant vector: $\vec V = \begin{pmatrix}
v_1 \\
v_2 \\
v_3 \\
\end{pmatrix}$, $\Lambda$ is a c-number.

Thus, the general solution of the non-homogeneous system of equations
can be written as ({Concerning some exclusions,} %MDPI: Footnote is not permitted in this journal, so we have moved it into the text, please confirm the whole text.
 see an important remark at the end of this subsection.)
\be \label{gensol} \begin{pmatrix}
f_{11} (t) \\
f_{12} (t) \\
f_{21} (t) \\
\end{pmatrix} = \begin{pmatrix}
f_{p11} \\
f_{p12} \\
f_{p21} \\
\end{pmatrix}+ c_1 e^{\Lambda_1 t} \begin{pmatrix}
v^{(1)}_1 \\
v^{(1)}_2 \\
v^{(1)}_3 \\
\end{pmatrix} +
c_2 e^{\Lambda_2 t} \begin{pmatrix}
v^{(2)}_1 \\
v^{(2)}_2 \\
v^{(2)}_3 \\
\end{pmatrix} +
c_3 e^{\Lambda_3 t} \begin{pmatrix}
v^{(3)}_1 \\
v^{(3)}_2 \\
v^{(3)}_3 \\
\end{pmatrix}
\ee
where $c_1, c_2, c_3$ are arbitrary complex constants.

Substituting (\ref{x}) into (\ref{homdiffeq}), one gets
\be \Lambda e^{\Lambda t} \vec V = A e^{\Lambda t} \vec V \Rightarrow \ee
\be \label{eig0} \Rightarrow A \vec V = \Lambda \vec V\ee

Thus, we need the eigenvalues and eigenvectors of the matrix $A$ (defined in (\ref{homdiffeq})).
Equation (\ref{eig0}) can be represented in the following form:
{\footnotesize
\ba
&&\!\!\!\!\!\!\!\!\begin{pmatrix}
0 & i e_{21} & -i e_{12} \\
2i e_{12}&-i \Delta e  - \frac{1}{2} |\lambda_1|^2 - \frac{1}{2} |\lambda_2|^2 + \lambda_1 \lambda^*_2 & 0 \\
-2i e_{21}& 0 & i\Delta e  - \frac{1}{2} |\lambda_1|^2 - \frac{1}{2} |\lambda_2|^2 + \lambda^*_1 \lambda_2 \\
\end{pmatrix}
\begin{pmatrix}
v_1 \\
v_2 \\
v_3 \\
\end{pmatrix} = \nonumber\\
&&= s
\begin{pmatrix}
v_1 \\
v_2 \\
v_3 \\
\end{pmatrix}, \label{eigdiag}
\ea
}
where we have made a new notation: $s = \frac{\Lambda}{c^2}, e_{ij} = \frac{\eps_{ij}}{c^2}$.

The characteristic polynomial of the matrix in Equation (\ref{eigdiag}) is
\ba &&s^3 + s^2 |\lambda_1 - \lambda_2|^2\nonumber\\&& + s \left[ 4 |e_{12}|^2 + \left|\Delta e + \frac{1}{2} i (\lambda_1 \lambda^*_2 - \lambda^*_1 \lambda_2)\right|^2 + \frac{1}{4} |\lambda_1 - \lambda_2|^4 \right] +\nonumber \\
&&+ 2 |e_{12}|^2 |\lambda_1 - \lambda_2|^2 = 0
\label{cubicdiag}
\ea

Next, we have to consider two cases: when the last term vanishes and when it does not. These two cases give completely different dynamics of a system.
\begin{enumerate}
\item $\bs{e_{12} = e_{21} = 0$ or $\lambda_1 = \lambda_2}$:\\
{Solving} %MDPI: is the bold necessary for the above equation? please confirm. If not, please remove.
 Equation (\ref{cubicdiag}), we obtain
\be
%\label{diagspec1}
s_1 = 0
%\nonumber
\ee
\be
%\label{diagspec2}
s_{2,3} = -\frac{1}{2} |\lambda_1 - \lambda_2|^2 \pm i \sqrt{4|e_{12}|^2 + \left|e_{11} - e_{22} + \frac{1}{2} i (\lambda_1 \lambda_2^* - \lambda_1^* \lambda_2)\right|^2}
%\nonumber
\ee

We see that, when $\lambda_1 \neq \lambda_2$, according to (\ref{gensol}), last two exponents decrease, while the first one reduces to a constant, which merges with a pointer. Therefore, we conclude that, during the evolution, such a system approaches a constant that is not constrained by interaction with an environment, parameters in Hamiltonian, etc.

When $\lambda_1 = \lambda_2$, the real part of $s_{2,3}$ vanishes, and we have neverending oscillations of a solution.
\item $\bs{|e_{12}|^2 |\lambda_1 - \lambda_2|^2 \neq 0}$:\\
Below, we will prove that, for this case $\on{Re} s < 0$ for each root $s$ of this equation, i.e, according to (\ref{gensol}), $\rho(t)$ converges to the pointer $\rho_p$ for $t\rightarrow \infty$. In general, two options are possible for this case.
\begin{enumerate}
\item All three roots of (\ref{cubicdiag}), $s_1, s_2, s_3$, are real \\
The form of l.h.s. of (\ref{cubicdiag}) is such that it is strictly positive for $s \geq 0$, and therefore, $s_1, s_2, s_3$ have to be negative.

\item Equation (\ref{cubicdiag}) has two complex roots, $s, s^*$, and one real root $t$ 

To start with, we write Vieta's formulas for Equation (\ref{cubicdiag}):
\be \label{sys1diag} 2 \on{Re} (s) + t = -|\lambda_1 - \lambda_2|^2 \ee
\be  \label{sys2diag} |s|^2 + 2 \on{Re} (s) t =4 |e_{12}|^2 + \left|\Delta e + \frac{1}{2} i (\lambda_1 \lambda^*_2 - \lambda^*_1 \lambda_2)\right|^2 + \frac{1}{4} |\lambda_1 - \lambda_2|^4 \ee
\be \label{sys3diag} |s|^2 t = - 2 |e_{12}|^2 |\lambda_1 - \lambda_2|^2 \ee

$|s|^2 \neq 0$, since $s=0$ cannot be the root of (\ref{cubicdiag}). Therefore, from the last equation, it is obvious that $t<0$.

We are left to prove that $\on{Re} (s) < 0$.
Expressing $|s|^2$ and $t$ from the system of Equations (\ref{sys1diag})--(\ref{sys3diag}), we get the equation for $\on{Re} s$:
\be    \on{Re} (s)^3 + |\lambda_1 - \lambda_2|^2 \on{Re} (s)^2
\nonumber
\ee
\be + \left( |e_{12}|^2 + \frac{1}{4} \left|\Delta e + \frac{1}{2} i (\lambda_1 \lambda^*_2 - \lambda^*_1 \lambda_2)\right|^2 + \frac{5}{16} |\lambda_1 - \lambda_2|^4  \right) \on{Re} (s) \nn\ee
\be + \left(\frac{1}{4} |e_{12}|^2 + \frac{1}{8}\left|\Delta e + \frac{1}{2} i (\lambda_1 \lambda^*_2 - \lambda^*_1 \lambda_2)\right|^2 + \frac{1}{32} |\lambda_1 - \lambda_2|^4 \right)|\lambda_1 - \lambda_2|^2 = 0
%\nn
\ee

As before, we notice that $\on{Re} s \geq 0$ cannot be a solution of this equation; therefore, $\on{Re} s < 0$.
\end{enumerate}
The fact that $\on{Re} s_1, \on{Re} s_2, \on{Re} s_3 < 0$ is very important. It signifies the vanishing of the exponents in the general solution of the Lindblad equation (\ref{gensol}) over time (according to the definition, $\Lambda_1 = c^2 s$, $\Lambda_2 = c^2 s^*$, $\Lambda_3 = c^2 t$), which, finally, gives that the density matrix for any values of the parameters in $H, L$ (of the form (\ref{ljordan})) converges to the pointer (\ref{pointer2}).

\end{enumerate}

The other forms of the solution, not of the type (\ref{gensol}), one can find in Appendix \ref{appb}, where we analyze the coinciding roots of the Equation (\ref{cubicdiag}).

\subsection{The Case of the Lindblad Operator of the Jordan Block Form}

For the case of the Lindblad operator in the Jordan form (\ref{ljordan}), the FGKLS equation (\ref{lindblad}) also can be rewritten in a matrix form:
\be \label{homdiffeq++} \begin{pmatrix}
\dot f_{11} \\
\dot f_{12} \\
\dot f_{21} \\
\end{pmatrix} = c^2
\begin{pmatrix}
-1 & i e_{21}+ \frac{1}{2} \lambda & -ie_{12} + \frac{1}{2} \lambda^* \\
2i e_{12} - \lambda^* & -i(e_{11}-e_{22}) - \frac{1}{2} & 0 \\
-2ie_{21} - \lambda & 0 & i(e_{11} - e_{22}) - \frac{1}{2} \\
\end{pmatrix}
\begin{pmatrix}
f_{11} \\
f_{12} \\
f_{21} \\
\end{pmatrix}
\ee
and $f_{22}$ can be found from the trace condition.
Equation (\ref{eig0}) can be represented in the following form:
\be \label{eig}\begin{pmatrix}
-1 & i e_{21}+ \frac{1}{2} \lambda & -ie_{12} + \frac{1}{2} \lambda^* \\
2i e_{12} - \lambda^* & -i(e_{11}-e_{22}) - \frac{1}{2} & 0 \\
-2ie_{21} - \lambda & 0 & i(e_{11} - e_{22}) - \frac{1}{2} \\
\end{pmatrix}
\begin{pmatrix}
v_1 \\
v_2 \\
v_3 \\
\end{pmatrix} = s
\begin{pmatrix}
v_1 \\
v_2 \\
v_3 \\
\end{pmatrix} \ee
where $s$ is defined as follows: $s = \frac{\Lambda}{c^2}$.

The characteristic polynomial of the matrix in Equation (\ref{eig}) is
\ba&& \label{cubic} s^3 + 2s^2 + \left( \frac{5}{4} + (e_{11}-e_{22})^2 + 4 \left| \frac{1}{2}\lambda + ie_{21}\right|^2\right) s\nonumber\\&& + \left( \frac{1}{4} + (e_{11}-e_{22})^2 + 2\left|\frac{1}{2}\lambda + ie_{21}\right|^2 \right) = 0 \ea

Next, we have to prove that $\on{Re} s < 0$ for each root $s$ of this equation. It means, according to (\ref{gensol}), that $\rho(t)$ converges to the pointer $\rho_p$ for $t\rightarrow \infty$.

Two options exist for the cubic equation (\ref{cubic}):
\begin{enumerate}
\item If the roots of (\ref{cubic}), $s_1, s_2, s_3$ are real, they are negative. Otherwise, the left part of (\ref{cubic}) would be strictly positive but not $0.$
\item If Equation (\ref{cubic}) has two complex roots $s, s^*$, and one real root $t,$ all of them have negative real parts. This important fact is derived by means of the well-known \mbox{Vieta's formulas.}

To start with, we write Vieta's formulas for Equation (\ref{cubic}):
\be \label{sys1} 2 \on{Re} (s) + t = -2 \ee
\be  \label{sys2} |s|^2 + 2 \on{Re} (s) t =\frac{5}{4} + (e_{11}-e_{22})^2 + 4 \left| \frac{1}{2}\lambda + ie_{21}\right|^2 \ee
\be \label{sys3} |s|^2 t = - \left(\frac{1}{4} + (e_{11}-e_{22})^2 + 2\left|\frac{1}{2}\lambda + ie_{21}\right|^2 \right) \ee

$|s|^2 \neq 0$, since $s=0$ cannot be the root of (\ref{cubic}). Therefore, from the last equation, it is obvious that $t<0$.

We are left to prove that $\on{Re} (s) < 0$.
Expressing $|s|^2$ and $t$ from the system of \mbox{Equations (\ref{sys1})--(\ref{sys3}),} we get the cubic equation for $\on{Re} s$:
\be  \on{Re} (s)^3 +  2 \on{Re} (s)^2 + \on{Re} (s) \left(\frac{21}{16} + \frac{1}{4} (e_{11} - e_{22})^2 +\left|\frac{1}{2}\lambda + ie_{21}\right|^2\right) + \nn\ee
\be \left(\frac{9}{32} + \frac{1}{8}(e_{11} - e_{22})^2 + \frac{3}{4}\left|\frac{1}{2}\lambda + ie_{21}\right|^2 \right)= 0.
%\nonumber
\ee

As before, we conclude that $\on{Re} s \geq 0$ cannot be a solution of this equation; therefore, $\on{Re} s < 0$.

\end{enumerate}

The fact that $\on{Re} s_1, \on{Re} s_2, \on{Re} s_3 < 0$ is very important. It signifies the vanishing of the exponents in the general solution of the FGKLS equation (\ref{gensol}) over time (according to the definition, $\Lambda_1 = c^2 s$, $\Lambda_2 = c^2 s^*$, $\Lambda_3 = c^2 t$), which, finally, gives that the density matrix for any values of the parameters in $H, L$ (of the form (\ref{ljordan})) converges to the pointer (\ref{pointer2}).

%As in the previous Subsection, the case of coinciding roots must be considered separately.

For the Linblad operator of the Jordan block form, we have also found the other forms of the solution, not of the type (\ref{gensol}), corresponding to coinciding roots of (\ref{cubic}). One can find the complete analysis in Appendix \ref{appb}.

\section{Positivity of the Solution of FGKLS Equation}

Next, we have to make sure that the remaining property of the density matrix $\rho(t)$ is satisfied---namely, that it is positive. Otherwise, it does not have physical meaning. For this purpose, we shall take explicitly into account in (\ref{gensol}) that the condition $\Tr \rho(t) = 1$ (\ref{trace2}) has to be satisfied for any moment of time including asymptotical $t\to\infty .$
Therefore, the solution~(\ref{gensol}) can be rewritten in a matrix form (we consider only the case of non-coinciding roots $s_i$):
\be \label{rhot0} \rho(t) =
\begin{pmatrix}
f_{p11} & f_{p12} \\
f_{p21} & f_{p22} \\
\end{pmatrix} + \nonumber\ee
\be c_1 e^{s_1 c^2 t}
\begin{pmatrix}
v^{(1)}_1 & v^{(1)}_2 \\
v^{(1)}_3 & -v^{(1)}_1 \\
\end{pmatrix} + c_2 e^{s_2 c^2 t}
\begin{pmatrix}
v^{(2)}_1 & v^{(2)}_2 \\
v^{(2)}_3 & -v^{(2)}_1 \\
\end{pmatrix}+ c_3 e^{s_3 c^2 t}
\begin{pmatrix}
v^{(3)}_1 & v^{(3)}_2 \\
v^{(3)}_3 & -v^{(3)}_1 \\
\end{pmatrix}
\ee

After this, we shall parameterize it
conveniently, representing all the complex numbers in the polar form:
\be c_k = r_k e^{i \phi_k}, \;\;\;  v_l^{(k)} = \alpha_l^{(k)} e^{i \beta^{(k)}}, \;\;\;\;\; k,l = 1,2,3\ee
and then constrain the expression by the hermiticity condition (\ref{herm2}).
We obtain the following parametrization:
\begin{itemize}
\item For the case of two complex roots and one real root (${s_1,s_2 = s_1^* \in \mathbb{C}}, {s_3 \in \mathbb{R}}$):
\be \label{rhoparam1} \rho(t) = \begin{pmatrix}
f_{p11} & f_{p12} \\
f_{p21} & f_{p22} \\
\end{pmatrix} + u e^{-i \phi} e^{s_1 c^2 t} \begin{pmatrix}
d e^{-i\delta} & b e^{-i\gamma} \\
a e^{-i\alpha} & -d e^{-i\delta}
\end{pmatrix}\nonumber\ee
\be +
u e^{i \phi} e^{s^*_1 c^2 t} \begin{pmatrix}
d e^{i\delta} & a e^{i\alpha} \\
b e^{i\gamma} & -d e^{i\delta}
\end{pmatrix}  \pm
 w e^{s_3 c^2 t} \begin{pmatrix}
h & p e^{i \frac{\beta}{2}} \\
p e^{-i \frac{\beta}{2}} & -h \end{pmatrix} \ee
where $a,b,d,p,h,u,w\in \mathbb{R}_+$, $\alpha,\beta,\gamma,\delta, \phi \in \mathbb{R}$.\\
Since $(v^{k}_1,v^{k}_2,v^{k}_3)$ is an eigenvector of the matrix $A$ from (\ref{homdiffeq}), $\{v_l^{(k)}\}$ are completely determined by the particular form of the Hamiltonian $H$ and Lindblad operator $L$, i.e., by the values of parameters $\eps_1, \eps_2, c, \lambda$. Therefore, $\{a,b,d,p,h,\alpha,\beta,\gamma,\delta\}$ are also completely determined by $H$ and $L$. As for $\{u,w\in R_+,\phi\in R$,''$\pm$'' sign$\}$, these are free parameters. They depend on the choice of the initial state $\rho(0)$.
\item For the case of three real roots ($s_1, s_2, s_3 \in \mathbb{R}$):
\be \label{rhoparam2} \rho(t) = \begin{pmatrix}
f_{p11} & f_{p12} \\
f_{p21} & f_{p22} \\
\end{pmatrix}  \pm \nonumber\ee
\be
  w^{(1)} e^{s_1 c^2 t} \begin{pmatrix}
h^{(1)} & p^{(1)} e^{i \frac{\beta^{(1)}}{2}} \\
p^{(1)} e^{-i \frac{\beta^{(1)}}{2}} & -h^{(1)} \end{pmatrix}
 \pm w^{(2)} e^{s_2 c^2 t} \begin{pmatrix}
h^{(2)} & p^{(2)} e^{i \frac{\beta^{(2)}}{2}} \\
p^{(2)} e^{-i \frac{\beta^{(2)}}{2}} & -h^{(2)} \end{pmatrix} \pm \nonumber\ee
\be  w^{(3)} e^{s_3 c^2 t} \begin{pmatrix}
h^{(3)} & p^{(3)} e^{i \frac{\beta^{(3)}}{2}} \\
p^{(3)} e^{-i \frac{\beta^{(3)}}{2}} & -h^{(3)} \end{pmatrix}  \ee

where $\{ p^{(k)}, h^{(k)} \in \mathbb{R}_+, \beta^{(k)} \in \mathbb{R}, \; k=1,2,3 \}$ are determined by $H$ and $L$, while $\{\pm$ signs, $w^{(k)} \in \mathbb{R}_+, \; k=1,2,3\}$ depend on the choice of the initial state $\rho(0)$.
\end{itemize}

Interestingly, the initial state has just exactly three real parameters: $x,y,z \in \mathbb{R}$,
\be \rho (0) = \begin{pmatrix}
z & x+ iy \\
x-iy & 1-z \\
\end{pmatrix}
\nonumber
\ee

The correspondence between $\{x,y,z \in \mathbb{R}\}$ and $\{u,w \in \mathbb{R}_+;\phi \in \mathbb{R};\pm\}$ or  $\{\pm$ signs, $w^{(k)} \in \mathbb{R}_+, \; k=1,2,3\}$ is a problem to be solved. The choice of signs ''$\pm$'' could signify several different paths that the system may choose to take.

In order to check the positivity of (\ref{rhoparam1}) and (\ref{rhoparam2}), we need to take the determinant of this matrix $\rho(t)$ and find the conditions (i.e., the moments of time $t$), when the determinant $\det{\rho(t)}$ is non-negative. The resulting inequalities are not solvable explicitly in a general form.
Therefore, we will restrict ourselves by the particular case when there is only the last exponent in (\ref{rhoparam1}) and (\ref{rhoparam2}) with a real value $s_3:$
\be \label{redsol}\rho(t) =
\begin{pmatrix}
f_{p11} & f_{p12} \\
f_{p21} & f_{p22} \\
\end{pmatrix} \pm w e^{s_3 c^2 t}
\begin{pmatrix}
h  & p e^{i \frac{\beta}{2}} \\
p e^{-i \frac{\beta}{2}} & -h  \\
\end{pmatrix}
\ee

We will denote $\pm w e^{s_3 c^2 t}$ as $x$, and the positivity of the density matrix means:
\be \det{\rho(t)} = \left| \begin{matrix}
f_{p11} + x h  & f_{p12} + x p e^{i \frac{\beta}{2}} \\
f_{p21} + x p e^{-i \frac{\beta}{2}} & f_{p22} - x h \\
\end{matrix} \right| \nonumber\ee
\be = - x^2 \left(p^2 +h^2\right) + x \left(h f_{p22} - h f_{p11} - p e^{i \frac{\beta}{2}} f_{p21} - p e^{-i \frac{\beta}{2}} f_{p12}\right) + f_{p11} f_{p22} - f_{p21} f_{p12}\nonumber\ee
\be = -\left(p^2+ h^2 \right) (x-x_1) (x-x_2) \geq 0,\ee
%for $\rho(t)$ to be positive.\\
or,
\be \label{ineqexp} x_1 \leq \pm w e^{s_3 c^2 t} \leq x_2 \ee
where $x_1, x_2, \;x_1 \leq x_2$, are the roots of the quadratic polynomial above.
%corresponding square equation.

One of Vieta's formulas gives
\be -(p^2+h^2) x_1 x_2 = f_{p11} f_{p22} - f_{p21} f_{p12} \geq 0
%\nonumber
\ee
{The} latter inequality follows from the positivity of the pointers in Section \ref{sec3}.
It means that $x_1 x_2 \leq 0 \Rightarrow x_1 \leq 0, x_2 \geq 0$.

There are three options to fulfill the inequality (\ref{ineqexp}):
\begin{itemize}
\item $w=0$\\
It is a trivial case. $\rho(t)$ is just a pointer in any moment of time $t$. It is positive, as we checked earlier, and the inequality (\ref{ineqexp}) is obviously correct.
\item ''$+$'' sign, $w>0$ \\
The inequalities (\ref{ineqexp}) are reduced to:
\be s_3 c^2 t \leq \ln{x_2} - \ln{w}.
%\nonumber
\ee
Thus, the solution $\rho(t)$ (\ref{redsol}) of the FGKLS equation has physical sense only for
\be t \geq \frac{\ln{w} - \ln{x_2}}{-s_3 c^2}
%\nn
\ee
\item ''$-$'' sign, $w>0$ \\
In a similar way, we obtain:
\be s_3 c^2 t \leq \ln{(-x_1)} - \ln{w},
%\nonumber
\ee
and possible time is:
\be t\geq \frac{ \ln{w} -\ln{(-x_1)}}{-s_3 c^2}.
%\nn
\ee
\end{itemize}
For $w\neq 0,$ only these time intervals provide a reasonable solution of the FGKLS equation.

\section{The Behavior of Solutions
for Weak Interaction with Environment}

In this section, we examine the behavior of the solution (\ref{rhot0}) when interaction with an environment disappears, i.e., for $c \rightarrow 0.$

We start from the Lindblad operator $L$ of diagonal form (\ref{ldiag}) solving
Equations (\ref{eigdiag}) and (\ref{cubicdiag}) by means of perturbation theory with parameter $c^2$. For simplicity, we take the case of diagonal Hamiltonian $\eps_{12} = \eps_{21} = 0$.
Since, in all the equations, we have integer powers of $c^2$, we decompose the parameters $\Lambda$ of the general solution (\ref{gensol}) in a series:  %over $c^2$. To compensate for the factor $c^2$ in $\Lambda = c^2 s$ (see (\ref{gensol}),(\ref{eig})), we start the expansion of $s$ from $\frac{1}{c^2}$:
\be \Lambda = c^2 s = a_0  + a_1 c^2 + \dots
%\nonumber
\ee

In the leading order, Equation (\ref{cubic}) reads:
\be a_0^3 + a_1 (\eps_{11}-\eps_{22})^2= 0
%\nonumber
\ee
with three solutions: $a_0 = 0,$ and $a_0 = \pm i (\eps_{11}-\eps_{22})$.
Considering next-to-leading order of Equation (\ref{cubicdiag}), we obtain for $a_1:$
\be a_1 = -\frac{|\lambda_1 - \lambda_2|^2 a_0^2  + a_0 i (\lambda_1 \lambda_2^* - \lambda_1^* \lambda_2)\Delta\eps}{3 a_0^2 + (\eps_{11}-\eps_{22})^2}
%\nonumber
\ee
which gives three options:
\begin{itemize}
\item $a_0 = 0$ and $a_1 = 0;$
\item $a_0 =  i (\eps_{11}-\eps_{22})$ and $a_1  = -\frac{1}{2}(|\lambda_1|^2 + |\lambda_2|^2 - 2 \lambda_1^* \lambda_2);$
\item $a_0 = - i (\eps_{11}-\eps_{22})$ and $a_1 = -\frac{1}{2}(|\lambda_1|^2 + |\lambda_2|^2 - 2 \lambda_1 \lambda_2^*).$
\end{itemize}

Then, the general solution of (\ref{gensol}) with diagonal $L$ for small $c^2$ looks like:
\be \rho(t) = v_{const} + e^{i\Delta\eps t -\frac{1}{2}(|\lambda_1|^2 + |\lambda_2|^2 - 2 \lambda_1^* \lambda_2) c^2 t + \dots} v_1+  e^{-i\Delta\eps t -\frac{1}{2}(|\lambda_1|^2 + |\lambda_2|^2 - 2 \lambda_1 \lambda_2^*) c^2 t + \dots} v_1^\dag
%\nonumber
\ee
where $v_{const}, v_1$ are time-independent matrices. One can check explicitly that the real parts in both exponents of this expression are negative.

For the case of the Lindblad operator with Jordan block structure (\ref{ljordan}) and diagonal Hamiltonian $H$ for small values of $c^2,$ the behavior of the solution (\ref{rhot0}) can be obtained in a similar way:
\be \Lambda = c^2 s = a_0 + a_1 c^2 + \dots
%\nonumber
\ee

In the leading order, Equation (\ref{cubic}) reads:
\be a_0^3 + a_0 (\eps_{11}-\eps_{22})^2 = 0,
%\nonumber
\ee
providing again three solutions: $a_0 = 0$ and $a_0 = \pm i (\eps_{11}-\eps_{22}).$
Next, we find $a_1$ considering next-to-leading order of Equation (\ref{cubic}):
\be a_1 = -\frac{2 a_0^2 + (\eps_{11}-\eps_{22})^2}{3 a_0^2 + (\eps_{11}-\eps_{22})^2}
%\nonumber
\ee
which gives:
\begin{itemize}
\item $a_0 = 0;\,\, a_1 = -1;$
\item $a_0 =  i (\eps_{11}-\eps_{22}); \,\, a_1 = -\frac{1}{2};$
\item $a_0 = - i (\eps_{11}-\eps_{22}); \,\, a_1 = -\frac{1}{2}$
\end{itemize}

Finally, solving the eigenvector Equation (\ref{eig}) in the leading and next-to-leading order, we get for the density matrix:
\ba
\rho(t) &=& \begin{pmatrix}
f^{(p)}_{11} & f^{(p)}_{12} \\
f^{(p)}_{21} & f^{(p)}_{22} \\
\end{pmatrix} + c_1 e^{- c^2 t + \dots}
\begin{pmatrix}
1 + x c^2 + \dots & i \frac{\lambda^* c^2}{\eps_{11} - \eps_{22}} +\dots \\
- i \frac{\lambda c^2}{\eps_{11} - \eps_{22}} +\dots  & -1 - x c^2 - \dots \\
\end{pmatrix} + \nonumber \\
& & c_2 e^{i\Delta\eps t - \frac{1}{2} c^2 t + \dots}
\begin{pmatrix}
-\frac{1}{2}  i \frac{\lambda^* c^2}{\eps_{11} - \eps_{22}} + \dots & 0 + \dots \\
1+ y c^2 + \dots & \frac{1}{2}  i \frac{\lambda^* c^2}{\eps_{11} - \eps_{22}} + \dots \\
\end{pmatrix}+\nonumber \\
& & c_3 e^{-i\Delta\eps t - \frac{1}{2} c^2 t + \dots}
\begin{pmatrix}
\frac{1}{2}  i \frac{\lambda c^2}{\eps_{11} - \eps_{22}} + \dots & 1+ z c^2 + \dots \\
0 + \dots & -\frac{1}{2}  i \frac{\lambda c^2}{\eps_{11} - \eps_{22}} + \dots \\
\end{pmatrix},
%\nonumber
\ea
where $c_1, c_2, c_3, x, y, z$ are arbitrary complex numbers.

Applying the hermiticity condition (\ref{herm2}), we further constrain the solution:
\ba
\rho(t) &=& \begin{pmatrix}
f^{(p)}_{11} & f^{(p)}_{12} \\
f^{(p)}_{21} & f^{(p)}_{22} \\
\end{pmatrix} +  p e^{- c^2 t + \dots}
\begin{pmatrix}
1 + a c^2 + \dots & i \frac{\lambda^* c^2}{\eps_{11} - \eps_{22}} +\dots \\
- i \frac{\lambda c^2}{\Delta\eps} +\dots  & -1 - a c^2 - \dots \\
\end{pmatrix} + \nonumber \\
& & q e^{i\Delta\eps t - \frac{1}{2} c^2 t + \dots}
\begin{pmatrix}
-\frac{1}{2}  i \frac{\lambda^* c^2}{\Delta\eps} + \dots& 0 + \dots \\
1+ y c^2 + \dots & \frac{1}{2}  i \frac{\lambda^* c^2}{\Delta\eps} + \dots \\
\end{pmatrix}+ \nonumber \\
& &q^* e^{-i\Delta\eps t - \frac{1}{2} c^2 t + \dots}
\begin{pmatrix}
\frac{1}{2}  i \frac{\lambda c^2}{\Delta\eps} + \dots & 1+ y^* c^2 + \dots \\
0 + \dots & -\frac{1}{2}  i \frac{\lambda c^2}{\Delta\eps} + \dots \\
\end{pmatrix}
%\nonumber
\ea
where $p, a \in \mathbb{R}$, $q, y \in \mathbb{C}$.

We see that if the Lindblad operator $L \neq 0$ has the form (\ref{ljordan}), it must be that
$$\rho(t) \rightarrow \begin{pmatrix}
f^{(p)}_{11} & f^{(p)}_{12} \\
f^{(p)}_{21} & f^{(p)}_{22} \\
\end{pmatrix},$$
when $t \rightarrow \infty$,
%(according to the \textbf{Theorem} \ref{res}).
i.e., this system has a pointer (the late-time state of the system) of a very specific form. This is the effect of decoherence---the loss of information in the process of evolution of an open system, when the final state loses information about the initial state.

However, if the Lindblad operator $L=0$, or $c=0$, then we obtain the ordinary solution of the Liouville equation with oscillating exponents. It contains arbitrary parameters, which can be fixed by initial conditions. In other words, it does not approach the Lindblad pointer
$\begin{pmatrix}
f^{(p)}_{11} & f^{(p)}_{12} \\
f^{(p)}_{21} & f^{(p)}_{22} \\
\end{pmatrix}$ in the late time limit. Evolution of the system fully depends on the initial state, and information about the initial state is contained in the final state.

We have seen the same situation in our previous work \cite{pertalg}, when (no matter how small) $L \neq 0$ produced a very specific form of the pointer. However, at the same time, $L=0$ suggested that the pointer is a matrix with arbitrary elements (if the Hamiltonian is non-degenerate, then it is a diagonal matrix with arbitrary elements; if the Hamiltonian is degenerate, then it is a matrix with arbitrary diagonal elements and elements, corresponding to \mbox{degenerate indices).}

\section{Unitons}

In this section, a special kind of open system---unitons---will be considered, such that, despite an openness, their density matrix evolution is unitary. In other words, the density matrix $\rho_u(t)$ of the system interacting with the environment still obeys the FGKLS equation with a single Lindblad operator $L$, but simultaneously its Lindbladian part vanishes:
\be \label{uniton1} L \rho_u L^\dag - \frac{1}{2} \left\{L^\dag L,\rho_u \right\} = 0\ee
\be \label{uniton2} \dot{\rho_u} = -i[H,\rho_u] \;\; \Leftrightarrow \;\; \rho_u (t) = e^{-i H t} \rho_u (0) e^{i H t}\ee
{Keeping} decompositions of $H$ and $L$ as in (\ref{h2}) {and} %MDPI: changed comma to `and', please confirm.
 (\ref{l2}), the uniton's density matrix is now:
\be \rho_u = \sum_{ij} f^{(u)}_{ij} \ket{\psi_i} \bra{\psi_k}.
\ee

The first relation (\ref{uniton1}) is written on this basis as follows:
\be
\sum_{k,l}l_{mk}f^{(u)}_{kl}l^*_{nl} - \frac{1}{2}\sum_{k,l}l^*_{km}l_{kl}f^{(u)}_{ln}-\frac{1}{2}\sum_{k,l}f^{(u)}_{mk}l^*_{lk}l_{ln} = 0.
%\nonumber
\ee
{It} can be transformed into an explicit form of a matrix with four indexes multiplied by a vector with two indexes:
\be \sum_{kl} A_{mn,kl} f^{(u)}_{kl} = 0,
%\nonumber
\ee
where
\be A_{mn,kl} \equiv l_{mk} l^*_{nl} - \frac{1}{2} \sum_s l^*_{sm} l_{sk} \delta_{ln} - \frac{1}{2} \sum_s \delta_{km} l^*_{sl} l_{sn}.
%\nonumber
\ee
{Thus}, the existence of unitons depends crucially on the determinant of this matrix.

\begin{itemize}
\item If the determinant does not vanish, the only solution is $f^{(u)}_{mn} = 0 \;\forall m,n$, and no unitons exist in this case.
\item If the determinant vanishes, three options appear:
\begin{itemize}
\item The solution $f^{(u)}_{mn}\}$ has one free parameter.\\
This parameter is eliminated by the trace condition $\sum_k f^{(u)}_{kk} =1$. We have a constant density matrix that has to satisfy (\ref{uniton2}). Thus, unitons exist only if they commute with Hamiltonian: $[H,\rho_u] = 0$. As a result, the only uniton does not depend on time
: $\rho_u (t) = \rho_u (0)$. Actually, it can be considered as a pointer.
\item The solution $\{u_{mn}\}$ has two parameters.\\
One of the parameters is eliminated by the trace condition ${\sum_k u_{kk} =1}$, and the dependence of the other on time is found from Equation (\ref{uniton2}).
We have a solution with no free parameters, depending on time in some way.
\item The solution $\{u_{mn}\}$ has three or more parameters.\\
The same, as in the previous case, but the solution, besides depending on time in some way, also contains one or more parameters. Their dependence on time can be chosen in any manner.
\end{itemize}
\end{itemize}

{We} %MDPI: add indentation, plese confirm.
 are not able to calculate the determinant for an arbitrary dimension of operators, and we shall consider again a simpler task of dimension two.
%}
Note that, for this problem, we do not need perturbation theory. Equation (\ref{uniton1}) loses its part related to the Hamiltonian, and all the terms there turn out to be of the same order.

Solving Equation (\ref{uniton1}) in two-dimensional Hilbert space is actually the same as finding pointers in two dimensions for a vanishing Hamiltonian. This means that we should look at our previous results of Section \ref{sec2} and take $\eps_{ij}=0$.
The matrix equation that we arrive at is Equation (\ref{matreq1}) with $J, H$ given by Formulas (\ref{jvar}) and (\ref{hvar}) with $\eps_{ij} = 0$.

As before, we consider two types of Lindblad operator:
\begin{enumerate}
\item Diagonal Lindblad operator (\ref{ldiag})

The uniton for this case is easily found from the previous results for the pointer (\ref{pointerdiag1}) and (\ref{pointerdiag2}):
\begin{enumerate}
\item $\lambda_1 = \lambda_2$, then\\
\be
\rho_u =
\begin{pmatrix}
f_{u11} & f_{u12} \\
f_{u21} & 1 - f_{u11}\\
\end{pmatrix}
%\nonumber
\ee
If this condition for $L$ is satisfied, then every density matrix turns out to be a uniton, obeying Equation (\ref{uniton2}).

\item $\lambda_1 \neq \lambda_2$, then\\
\be
%\label{constraint}
\rho_u =
\begin{pmatrix}
f_{u11} & 0 \\
0 & 1 - f_{u11}\\
\end{pmatrix}
%\nonumber
\ee
Solving (\ref{uniton2}) for the diagonal density matrix, we see that the density matrix reduces to a constant matrix. It is not possible to make non-trivial time dependence with Equation (\ref{uniton2}). Unitons do not exist for this case.
\end{enumerate}
\item Lindblad operator of the Jordan block form (\ref{ljordan})

The uniton is found from the previous result for the pointer (\ref{pointer2}):
\be
%\label{unitonjordan}
\rho_u =
\begin{pmatrix}
\frac{|\lambda|^2 + 1}{2|\lambda|^2 + 1} & -\frac{\lambda^*}{2|\lambda|^2 + 1}\\
-\frac{\lambda}{2|\lambda|^2 + 1} & \frac{|\lambda|^2}{2|\lambda|^2 + 1}\\
\end{pmatrix}
%\nonumber
\ee
Since this solution of Equation (\ref{uniton1}) does not contain any free parameters, we cannot construct a uniton that non-trivially evolves in time according to (\ref{uniton2}). Unitons do not exist for this case.
\end{enumerate}

{We} %MDPI:added indentation, please confirm
 conclude that the only case when unitons exist is when the Lindblad operator is proportional to the identity matrix. It just falls off from the FGKLS equation. It is a trivial case that can be reduced to the Liouville equation, for which there is no interaction with an environment. Therefore, the FGKLS equation does not have oscillating Liouville solutions for the Hilbert space of dimension $2$.

\section{Conclusions}

Throughout the paper, we have exhaustively studied the evolution of an open quantum system for a Hilbert space of dimension $2$. We obtained final fixed states of an evolution of an open system (called pointers), and then we found a solution to the FGKLS equation and proved that it always converges to a pointer. It signifies a decoherence process, when information about an initial state becomes lost during an evolution as a result of an interaction with an environment. After this, we checked that the solution has a physical meaning, i.e., the density matrix is Hermitian, positive and has trace equal to $1$, and found a moment of time starting from which the density matrix is positive, i.e., a Lindblad equation can be used. Next, we studied a behavior of the solution when an interaction with an environment is weak. When the interaction is on, the general solution of the FGKLS equation has a special type, leading over time to a pointer. When it is off, the solution is a standard oscillating one, provided by the Liouville equation. Finally, we found that the FGKLS equation does not have oscillating solutions, of the same form as the solutions of Liouville equation, for the Hilbert space of dimension $2$.

%\section{Appendix\remove[EI]{: Lindblad operator of a Jordan block form: an example}}

%%%%%%%%%%%%%%%%%%%%%%%%%%%%%%%%%%%%%%%%%%
\vspace{6pt}

\newpage

\section*{Appendix A. Expression for a Pointer for Weak Interaction with Environment}\label{appa}
%\add[EI]{\protect \section*{Appendix A: E}

For the special case of diagonal but non-degenerate Hamiltonian $\eps_1 \neq \eps_2, \eps_{12}=\eps_{21}=0$ and Lindblad operator of the Jordan block form (\ref{ljordan}), a pointer (\ref{pointer2}) can be decomposed in a series over the parameter $c$, if we imagine that this parameter is small (weak interaction with an environment):
\vspace{6pt}
\ba
&&f^{(p)}_{11} = 1 + \frac{|\lambda|^2}{4} \sum_{k=1}^{\infty} c^{4k} \frac{(-1)^k (\frac{1}{2}|\lambda|^2 + \frac{1}{4})^{k-1}}{(\eps_1-\eps_2)^{2k}} \label{fp11row}\\
&&f^{(p)}_{22} = \frac{|\lambda|^2}{4} \sum_{k=1}^{\infty} c^{4k} \frac{(-1)^{k+1} (\frac{1}{2}|\lambda|^2 + \frac{1}{4})^{k-1}}{(\eps_1-\eps_2)^{2k}} \\
&&f^{(p)}_{12} = \frac{1}{2} i \lambda^* \sum_{k=1; k \;\text{is odd}}^{\infty} c^{2k} \frac{(-1)^{\frac{1}{2}(k-1)}(\frac{1}{2} |\lambda|^2 +\frac{1}{4})^{\frac{1}{2}(k-1)}}{(\eps_1 - \eps_2)^k} + \nonumber\\
&& \;\;\;\;\;\;\;\;\;\;\; \frac{1}{4} \lambda^* \sum_{k=2; k \;\text{is even}}^{\infty} c^{2k} \frac{(-1)^{\frac{1}{2}k}(\frac{1}{2} |\lambda|^2 +\frac{1}{4})^{\frac{1}{2}k-1}}{(\eps_1 - \eps_2)^k}\\
&&f^{(p)}_{21} = -\frac{1}{2} i \lambda \sum_{k=1; k \;\text{is odd}}^{\infty} c^{2k} \frac{(-1)^{\frac{1}{2}(k-1)}(\frac{1}{2} |\lambda|^2 +\frac{1}{4})^{\frac{1}{2}(k-1)}}{(\eps_1 - \eps_2)^k} + \label{fp21row}\nonumber\\
&& \;\;\;\;\;\;\;\;\;\;\; \frac{1}{4} \lambda \sum_{k=2; k \;\text{is even}}^{\infty} c^{2k} \frac{(-1)^{\frac{1}{2}k}(\frac{1}{2} |\lambda|^2 +\frac{1}{4})^{\frac{1}{2}k-1}}{(\eps_1 - \eps_2)^k}
\ea
or, more explicitly,
\ba
&&f^{(p)}_{11} = 1 - c^4 \frac{|\lambda|^2}{4(\eps_1-\eps_2)^2} + c^8 \frac{|\lambda|^2 (\frac{1}{2} |\lambda|^2 + \frac{1}{4})}{4(\eps_1 - \eps_2)^4} + \dots
%\nonumber
 \\
&&f^{(p)}_{22} =  c^4 \frac{|\lambda|^2}{4(\eps_1-\eps_2)^2} - c^8 \frac{|\lambda|^2 (\frac{1}{2} |\lambda|^2 + \frac{1}{4})}{4(\eps_1 - \eps_2)^4} + \dots
%\nonumber
\\
&&f^{(p)}_{12} = c^2 \frac{i \lambda^*}{2(\eps_1 - \eps_2)} - c^4 \frac{\lambda^*}{4(\eps_1 - \eps_2)^2} -  \nonumber\\ && \;\;\;\;\;\;\;\;\;\;\; c^6 \frac{i\lambda^*(\frac{1}{2}|\lambda|^2+\frac{1}{4})}{2(\eps_1 - \eps_2)^3} + c^8 \frac{\lambda^*(\frac{1}{2}|\lambda|^2+\frac{1}{4})}{4(\eps_1-\eps_2)^4} + \dots
%\nonumber
\\
&&f^{(p)}_{21} = - c^2 \frac{i \lambda}{2(\eps_1 - \eps_2)} - c^4 \frac{\lambda}{4(\eps_1 - \eps_2)^2} + \nonumber\\ && \;\;\;\;\;\;\;\;\;\;\;  c^6 \frac{i\lambda(\frac{1}{2}|\lambda|^2+\frac{1}{4})}{2(\eps_1 - \eps_2)^3} + c^8 \frac{\lambda(\frac{1}{2}|\lambda|^2+\frac{1}{4})}{4(\eps_1-\eps_2)^4} + \dots
%\nonumber
\ea

%We see that, interestingly, the terms of the order $c^2, c^6$, etc. (that is $\sim l^2, l^6$, etc.) are absent in the expansions of $f_{p11}$ and $f_{p22}$.
%The same expansion series (\ref{fp11row})--(\ref{fp21row}) can be obtained by applying perturbation theory schemes, studied in our previous paper \cite{pertalg}.}

\section*{Appendix B. Other Forms of the Solution of the FGKLS Equation}\label{appb}

%If the characteristic polynomial ((\ref{cubicdiag}) or (\ref{cubic})) of the set of FGKLS differential equations has coinciding roots, then the form of the solution \textbf{is not} that of (\ref{gensol}). In this section we investigate this particular case.

{The construction of the general solution described above has some exclusions, which correspond to the coinciding roots of Equation~(\ref{cubicdiag}) or (\ref{cubic}). In this case, the \mbox{solution
(\ref{gensol})} has to be modified: the exponents, corresponding to coinciding roots, acquire} \mbox{polynomial multipliers.}

\begin{itemize}
\item \textbf{{Diagonal} %MDPI: is the bold necessary?.
 Lindblad operator}

The analysis of Equation (\ref{cubicdiag}) shows that the roots coincide when
\begin{enumerate}
\item $|e_{12}|^2 = \frac{1}{54} |\lambda_1 - \lambda_2|^4,$\\
and\\
$\left|\Delta e + \frac{1}{2} i (\lambda_1 \lambda^*_2 - \lambda^*_1 \lambda_2)\right|^2 = \frac{1}{108} |\lambda_1 - \lambda_2|^4$,\\
then
\be s_1 = s_2 = s_3 = -\frac{|\lambda_1 - \lambda_2|^2}{3}
\nonumber
\ee
The solution is of the form:
\be \rho(t) = \rho_p + e^{-\frac{|\lambda_1 - \lambda_2|^2}{3}  c^2 t} v_1 + t  e^{-\frac{|\lambda_1 - \lambda_2|^2}{3}  c^2 t} v_2 + t^2  e^{-\frac{|\lambda_1 - \lambda_2|^2}{3}  c^2 t} v_3,
%\nonumber
\ee
where $v_1, v_2, v_3$ are time-independent matrices.

\item Provided\\
$|\lambda_1 - \lambda_2|^4 > 48|e_{12}|^2 + 12 \left|\Delta e + \frac{1}{2} i (\lambda_1 \lambda^*_2 - \lambda^*_1 \lambda_2)\right|^2$\\
and\\
$\left[ \frac{1}{4} |\lambda_1 - \lambda_2|^4 - 12 |e_{12}|^2 - 3 \left|\Delta e + \frac{1}{2} i (\lambda_1 \lambda^*_2 - \lambda^*_1 \lambda_2)\right|^2 \right]^3 = $\\
$= |\lambda_1 - \lambda_2|^4 \left[ -\frac{1}{8} |\lambda_1 - \lambda_2|^4 + 9 |e_{12}|^2 - \frac{9}{2} \left|\Delta e + \frac{1}{2} i (\lambda_1 \lambda^*_2 - \lambda^*_1 \lambda_2)\right|^2  \right]^2$,

\begin{enumerate}
\item If \\
$|\lambda_1 - \lambda_2|^2 \Big[ -\frac{1}{8} |\lambda_1 - \lambda_2|^4 + 9 |e_{12}|^2$ \\$- \frac{9}{2} \left|\Delta e + \frac{1}{2} i (\lambda_1 \lambda^*_2 - \lambda^*_1 \lambda_2)\right|^2  \Big] > 0$\\
then
\ba &&s_1 = s_2 = -\frac{|\lambda_1 - \lambda_2|^2}{3}
\nonumber
\\&& + \frac{1}{3} \sqrt{\frac{1}{4} |\lambda_1 - \lambda_2|^4 - 12 |e_{12}|^2 - 3 \left|\Delta e + \frac{1}{2} i (\lambda_1 \lambda^*_2 - \lambda^*_1 \lambda_2)\right|^2} \nonumber\ea
\ba && s_3 = -\frac{|\lambda_1 - \lambda_2|^2}{3} \nonumber\\&&- \frac{2}{3} \sqrt{\frac{1}{4} |\lambda_1 - \lambda_2|^4 - 12 |e_{12}|^2 - 3 \left|\Delta e + \frac{1}{2} i (\lambda_1 \lambda^*_2 - \lambda^*_1 \lambda_2)\right|^2} \nonumber\ea
\item If\\
$|\lambda_1 - \lambda_2|^2 \Big[ -\frac{1}{8} |\lambda_1 - \lambda_2|^4 + 9 |e_{12}|^2 $ \\$- \frac{9}{2} \left|\Delta e + \frac{1}{2} i (\lambda_1 \lambda^*_2 - \lambda^*_1 \lambda_2)\right|^2  \Big] < 0$,\\
then
 $s_1 = s_2 = -\frac{|\lambda_1 - \lambda_2|^2}{3}$\\$ - \frac{1}{3} \sqrt{\frac{1}{4} |\lambda_1 - \lambda_2|^4 - 12 |e_{12}|^2 - 3 \left|\Delta e + \frac{1}{2} i (\lambda_1 \lambda^*_2 - \lambda^*_1 \lambda_2)\right|^2}$\\
 $s_3 = -\frac{|\lambda_1 - \lambda_2|^2}{3} $\\$+ \frac{2}{3} \sqrt{\frac{1}{4} |\lambda_1 - \lambda_2|^4 - 12 |e_{12}|^2 - 3 \left|\Delta e + \frac{1}{2} i (\lambda_1 \lambda^*_2 - \lambda^*_1 \lambda_2)\right|^2} $
\end{enumerate}
The solution is of the form:
\be \rho(t) = \rho_p + e^{s_1 c^2 t} \tilde v_1 + t  e^{s_1 c^2 t} \tilde v_2 +  e^{s_3 c^2 t} \tilde v_3 \ee
where $\tilde v_1, \tilde v_2, \tilde v_3$ are time-independent matrices.

\end{enumerate}

\item \textbf{{Lindblad} operator of a Jordan block form}

%As in the previous Subsection, we will consider separately the case of coinciding roots of the cubic equation (\ref{cubic}), since then the solution (\ref{gensol}) is modified. Namely, the exponents, corresponding to coinciding roots, acquire a polynomial factors.

In the following, we will use the notation: \\
$E\equiv (e_{11} - e_{22})^2$, $K \equiv \left(\frac{1}{2} \lambda + i e_{21}\right)\left(\frac{1}{2} \lambda^* - i e_{12}\right)$.

\begin{enumerate}

\item The analysis of Equation (\ref{cubic}) shows that all three roots coincide
\be s_1 = s_2 = s_3 = -\frac{2}{3}
\nonumber
\ee
when $K = \frac{1}{54}, E = \frac{1}{108}$.
Moreover, the solution is of the form:
\be \rho(t) = \rho_p + e^{-\frac{2}{3} c^2 t} v_1 + t  e^{-\frac{2}{3} c^2 t} v_2 + t^2  e^{-\frac{2}{3} c^2 t} v_3
%\nonumber
\ee
where $v_1, v_2, v_3$ are time-independent matrices.

\item Two of three roots coincide if
$E + 4K < \frac{1}{12}$ and $\left(\frac{1}{4} - 3 E - 12 K\right)^3 =\linebreak \left( \frac{1}{8} + \frac{9}{2} E - 9 K\right)^2$.

Then, two situations are possible:
\begin{enumerate}
\item $\frac{1}{36} + E - 2 K > 0$, then

\be s_1 = s_2 = -\frac{2}{3} + \frac{1}{3} \sqrt{\frac{1}{4} - 3E -12K}
\nonumber
\ee
\be s_3 = -\frac{2}{3} - \frac{2}{3} \sqrt{\frac{1}{4} - 3E -12K}
\nonumber
\ee

\item $\frac{1}{36} + E - 2 K < 0$, then
\be s_1 = s_2 = -\frac{2}{3} - \frac{1}{3} \sqrt{\frac{1}{4} - 3E -12K}
\nonumber
\ee
\be s_3 = -\frac{2}{3} + \frac{2}{3} \sqrt{\frac{1}{4} - 3E -12K}
\nonumber
\ee

\end{enumerate}

In both situations, the solution has the form:
\be \rho(t) = \rho_p + e^{s_1 c^2 t} \tilde v_1 + t  e^{s_1 c^2 t} \tilde v_2 +  e^{s_3 c^2 t} \tilde v_3 \ee
where $\tilde v_1, \tilde v_2, \tilde v_3$ are time-independent matrices.

\end{enumerate}

\end{itemize}

\section*{Appendix C. Lindblad Operator of a Jordan Block Form: An Example}\label{appc}

Provided that the current understanding of the quantum theory remains to be valid on the fundamental level, the Lindblad equation should arise as an effective description of a system as a subsystem of a larger system that includes environmental degrees of freedom that is assumed to undergo the unitary evolution driven by a self-adjoint Hamiltonian,
\begin{equation}
H_{full}=H_S\otimes I+I\otimes H_{env}+H_{int}.
%\nonumber
\end{equation}

The interaction Hamiltonian $H$ can always be decomposed into a superposition of the tensor products of the operators acting solely on the system and environment,
\begin{equation}
H_{int}=\sum_{k} A_{S,k}\otimes B_{k}.
%\nonumber
\end{equation}

To describe all the measurements performed on the subsystem of interest without affecting the environmental degrees of freedom, the reduced density matrix can be used:
\begin{equation}
\rho_S=\Tr_{env}\rho .
%\nonumber
\end{equation}

To derive the Lindblad equation, one has to apply certain approximations \cite{breuer} (p. 130).
\begin{enumerate}
\item \textit{{Born} %MDPI: is the italic necessary?..
 approximation}: we assume that, due to a weak coupling between system and environment, the total density matrix is close to a factorized one:
\begin{equation}
\rho\simeq \rho_S\otimes \rho_{env}.
%\nonumber
\end{equation}
The environmental density matrix is usually assumed to be stationary.
\item \textit{{Born--Markov} approximation}: {we assume that the correlation functions in} \mbox{the environment},
\begin{equation}
\langle B_i^\dagger(t) B_j(\tau)\rangle_{env}=\Tr_{env}\Big(\rho_{env}B_i^\dagger(t)B_j(\tau)\Big),
%\nonumber
\end{equation}
where time-dependent operators $A_{S,k}$ and $B_k$ are defined in the interaction picture, and decay sufficiently fast compared to the relaxation time scale of the system to the equilibrium with environment. This means that for the typical values of $t$, we can apply the following approximation:
%\begin{equation}
\begin{align}
&\int_0^t d\tau \langle B_i^\dagger(t) B_j(\tau)\rangle_{env} A_{S,l}(t)A_{S,k}(\tau) \rho_S(\tau)\simeq
\nonumber\\
&\left[\int_0^{+\infty} d\tau \langle B_i^\dagger(\tau) B_j(0)\rangle_{env} A_{S,l}(t)A_{S,k}(t-\tau)\right]\rho_S(t)
\end{align}
\item \textit{{Rotating} %MDPI: is the italic necessary?.
 wave approximation}: we assume that the relaxation time scale of the system to the equilibrium with the environment is much smaller than the time scale of the system.
We introduce the basis of the system Hamiltonian eigenoperators defined as
\begin{equation}
[H_S,A(\omega)]=-\omega A(\omega),
%\nonumber
\end{equation}
so that we can take $A_{S,\omega}(t)=A(\omega)e^{-i\omega t}$.
In the rotating wave approximation, the terms with different $\omega$ correspond to the rapid oscillations that can be neglected on the time scales under consideration.
\end{enumerate}

Applying this to the qubit interacting with the environment, we can choose the eigenbasis of the free Hamiltonian:
\begin{equation}
H_S=\begin{pmatrix}E_\uparrow&0\\0&E_\downarrow\end{pmatrix},
%\nonumber
\end{equation}
{There} %MDPI: can add indentation?.
 is a single eigenoperator for each of the non-zero frequencies $\pm\omega=\pm\frac{E_\downarrow-E_\uparrow}{2}$,
\begin{equation}
    A(\omega)=\begin{pmatrix}0&1\\0&0\end{pmatrix},\quad A(-\omega)=\begin{pmatrix}0&0\\1&0\end{pmatrix}
    %\nonumber
\end{equation}
and two eigenoperators for zero frequency:
\begin{equation}
    A_\uparrow(0)=\begin{pmatrix}1&0\\0&0\end{pmatrix},\quad A_\downarrow(0)=\begin{pmatrix}0&0\\0&1\end{pmatrix}
    %\nonumber
\end{equation}

The rotating wave approximations enforce the following form of the non-unitary part of the master equation:
\begin{align}
\Gamma_{++}\Big(2A(\omega)\rho  A(\omega)^\dagger-\{\rho,A(\omega)^\dagger A(\omega)\}\Big)+\nonumber\\
\Gamma_{--}\Big(2A(-\omega)\rho A(-\omega)^\dagger-\{\rho,A(-\omega)^\dagger A(-\omega)\}\Big)+\nonumber\\
\sum_{a,b=\uparrow,\downarrow}\Gamma_{ab}\Big(2A_a(0)\rho A_b(0)^\dagger-\{\rho,A_b(0)^\dagger A_a(0)\}\Big)
%\nonumber
\end{align}
where the $\Gamma$-parameters are determined by the environmental correlation functions
\begin{align}
\Gamma_{++}=\int_0^\infty dt \cos{\omega t} \langle B_{+}^\dagger(t) B_{+}(0)\rangle_{env}, \nonumber \\
\Gamma_{--}=\int_0^\infty dt \cos{\omega t} \langle B_{-}^\dagger(t) B_{-}(0)\rangle_{env}\\
\Gamma_{ab}=\int_0^{+\infty}dt \langle B^\dagger_a(t)
B_b(0)\rangle_{env}.
\nonumber
\end{align}

If we require that this part would be equivalent to the Lindbladian with a single Lindblad operator,
\begin{equation}
L\rho L^\dagger - \frac{1}{2}L^\dagger L\rho - \frac{1}{2}\rho L^\dagger L
\nonumber
\end{equation}
there are three possible options:
\begin{enumerate}
    \item $\Gamma_{++}=\Gamma_{--}=0$, and matrix $\Gamma_{ab}$ is arbitrary. This is equivalent to the Lindbladian with
    \begin{equation}
        L=\begin{pmatrix}0&c\\\bar{c}&0\end{pmatrix}, \quad L=L^\dagger,
        \nonumber
    \end{equation}
    where the parameter $c$ is determined by the matrix $\Gamma_{ab}$.
    \item $\Gamma_{++}\neq 0,\Gamma_{--}=0$, the contribution of the matrix $\Gamma_{ab}$ vanishes. This is equivalent to the Lindbladian with
    \begin{equation}
        L=\begin{pmatrix}0&2c\\0&0\end{pmatrix},\quad |c|^2=\frac{\Gamma_{++}}{2}
        \nonumber
    \end{equation}
    \item $\Gamma_{--}\neq 0,\Gamma_{++}=0$, the contribution of the matrix $\Gamma_{ab}$ vanishes. This is equivalent to the Lindbladian with
    \begin{equation}
        L=\begin{pmatrix}0&0\\2c&0\end{pmatrix},\quad |c|^2=\frac{\Gamma_{--}}{2}.
        \nonumber
    \end{equation}
\end{enumerate}
{The} second option that results in the Jordan form of the Lindblad operator may be achieved, e.g., for the reservoir of oscillators in the squeezed vacuum state with $B_{+}=\sum_k c_k a_k^\dagger$, $B_{-}=\sum_k c_k^\ast a_k$, resulting in $\Gamma_{--}=0$ but $\Gamma_{++}\neq 0$ \cite{breuer} (p. 149). The less restricted form of the non-diagonalizable Lindblad operator may imply a deviation from the rotating \mbox{wave approximation.}

\end{document}